# Out-of-plane displacement of quantum color centers in monolayer *h*-BN

Zhao Tang,[1,2,†] Fanhao Jia,[3,4,†] Greis J. Kim-Reyes,[1,5] Yabei Wu,[6] James R. Chelikowsky,[2,7,8] and Peihong Zhang[1*]

[1]*Department of Physics, University at Buffalo, State University of New York, Buffalo, New York 14260, USA*
[2]*Center for Computational Materials, Oden Institute for Computational Engineering and Sciences, The University of Texas at Austin, Austin, Texas 78712, USA*
[3]*Department of Physics, School of Sciences, Hangzhou Dianzi University, Hangzhou 310018, China*
[4]*School of Materials Science and Engineering & International Centre of Quantum and Molecular Structures, Shanghai University, Shanghai 200444, China*
[5]*Department of Physics and Astronomy, State University of New York at New Paltz, New Paltz, New York 12561, USA*
[6]*Department of Materials Science and Engineering, Southern University of Science and Technology, Shenzhen, Guangdong 518055, China*
[7]*Department of Physics, The University of Texas at Austin, Austin, Texas 78712, USA*
[8]*McKetta Department of Chemical Engineering, The University of Texas at Austin, Austin, Texas 78712, USA*
[†]Z.T. and F.J. contributed equally to this work.

*E-mail: pzhang3@buffalo.edu

Color centers exhibiting deep-level states within the wide bandgap *h*-BN monolayer possess substantial potential for quantum applications. Uncovering precise geometric characteristics at the atomic scale is crucial for understanding defect performance. In this study, first-principles calculations were performed on the most extensively investigated $C_BV_N$ and $N_BV_N$ color centers in *h*-BN, focusing on the out-of-plane displacement and their specific impacts on electronic, vibrational, and emission properties. We demonstrate the competition between the $\sigma^*$-like antibonding state and the $\pi$-like bonding state, which determines the out-of-plane displacement. The overall effect of vibronic coupling on geometry is elucidated using a pseudo Jahn-Teller model. Local vibrational analysis reveals a series of distinct quasi-local phonon modes that could serve as fingerprints for experimental identification of specific point defects. The critical effects of out-of-plane displacement during the quantum emission process are carefully elucidated to answer the distinct observations in experiments, and these revelations are universal in quantum point defects in other layered materials.

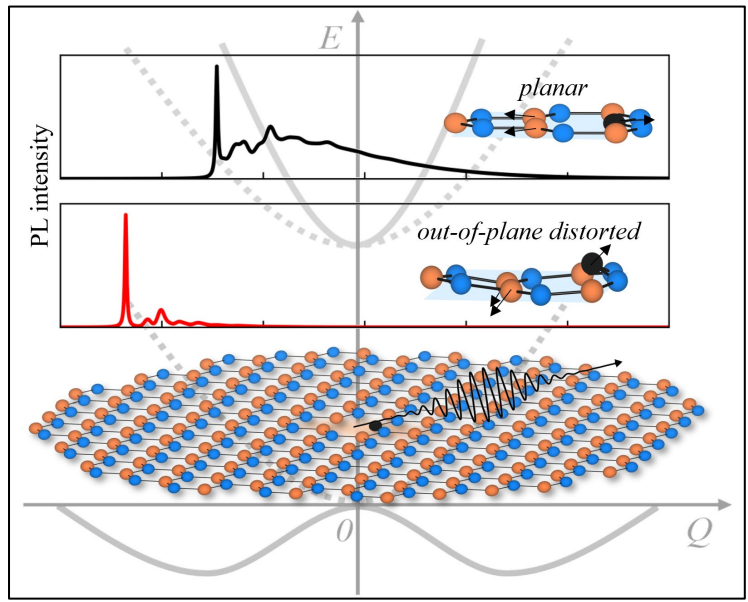

TOC figure

## INTRODUCTION

The discovery of polarized and ultrabright single-photon emitters (SPEs) in hexagonal boron nitride (*h*-BN) monolayers [1] has motivated extensive research efforts into quantum color centers in two-dimensional (2D) materials [2-5]. Certain color centers in h-BN exhibit exceptional properties such as high quantum efficiency and stable room-temperature functionality [6,7], making them promising candidates for practical applications like quantum sensing [8,9] and quantum information processing [10]. The prospect of site-specific fabrication methods offers a promising pathway to advance quantum technologies [11,12].

Point defect emitters in *h*-BN exhibit optical transition energies distributed across a large and continuous spectral range [1,9,13,14]. Within a range near 2 eV, individual experimental spectra display significant variability in the line shapes [15,16], which were categorized into two types, referred to as 'Group 1' and 'Group 2' [17]. Group 1 emitters have large phonon sideband (PSB) contributions, while Group 2 emitters have a strong zero-phonon line (ZPL) but weak PSB. The persistence of this distinction casts a shadow of uncertainty on the advancements of quantum defects in *h*-BN, underscoring the need for a precise understanding. Considering the challenges of site-specific characterization of point defects, first-principles simulations provide a crucial bridge between experimental emission spectra and specific chemical attributes of color centers [18-20].

Due to the strong coupling among structural, electronic, and vibrational properties, the quantum efficiency and performance of point defect emitters are highly sensitive to variations in their local structures. We notice that many previous theoretical simulations were based on the presumed fully planar defect structures [1,21-23], neglecting the out-of-plane degree of freedom. This assumption is inaccurate in most cases, as out-of-plane displacements can significantly alter the local symmetry and modify the electronic structure of defect sites, particularly for defect levels derived from out-of-plane $p_z$ orbitals. Indeed, Noh *et al.* found as early as 2018 that the optical transition activation energy can be changed by out-of-plane displacement [24]. Then, Li *et al.* [25] and Turiansky *et al.* [26] studied the manipulation of defect states and radiative lifetime by out-of-plane displacements. However, a comprehensive understanding of the out-of-plane displacements for specific defect types and their correspondence with measured data such as photoluminescence (PL) spectra remains lacking, which poses challenges for accurate identification and theoretical modeling.

To this end, we comparatively studied three defect candidate systems in monolayer *h*-BN: a spin singlet $C_BV_N$ (one carbon-for-boron substitution and one adjacent nitrogen vacancy), a spin triplet $C_BV_N$, and a spin doublet $N_BV_N$ (one nitrogen-for-boron substitution and one adjacent nitrogen vacancy), using density functional theory (DFT) calculations. Both $C_BV_N$ and $N_BV_N$ defects have been extensively studied over recent years, because their emission energies, polarizations, and radiative lifetimes align reasonably well with experimental observations [21,27]. Since the ground state of $C_BV_N$ is controversial, which may be either a spin triplet [22] or a nonmagnetic spin singlet [23], we investigated both states in this work. We aimed to reveal the origin of out-of-plane displacements by understanding the chemical nature of these three types of defects, particularly the underlying non-adiabatic coupling effects between electronic, vibrational properties and local displacements (both in-plane and out-of-plane displacements).

By evaluating total energies and dynamical stability, we confirmed that the ground state structures for spin singlet $C_BV_N$ and spin doublet $N_BV_N$ present *out-of-plane distortions* with $C_s$ symmetry, while the ground state structure of spin triplet $C_BV_N$ remains *planar* with a higher $C_{2v}$ symmetry. To unravel the mechanism behind these displacements, we studied the couplings between individual electronic states and the local displacements, showing that the occupation of $\sigma^*$-like antibonding states strongly correlates with out-of-plane instability, while $\pi$-like bonding states occupation helps stabilize planar structures. We introduced a two-state pseudo Jahn-Teller (PJT) model to analyze vibronic coupling on geometry. Our local vibrational analysis suggested a series of phonon modes distinct from those in pristine monolayer *h*-BN that can serve as fingerprints to help experimental identification of specific point defects. Finally, we quantified the critical impacts of out-of-plane displacement on vibronic structure during optical transitions using the Huang-Rhys approximation [28]. This analysis indicates that the distinction between 'Group 1' and 'Group 2' most likely originates from out-of-plane displacement.

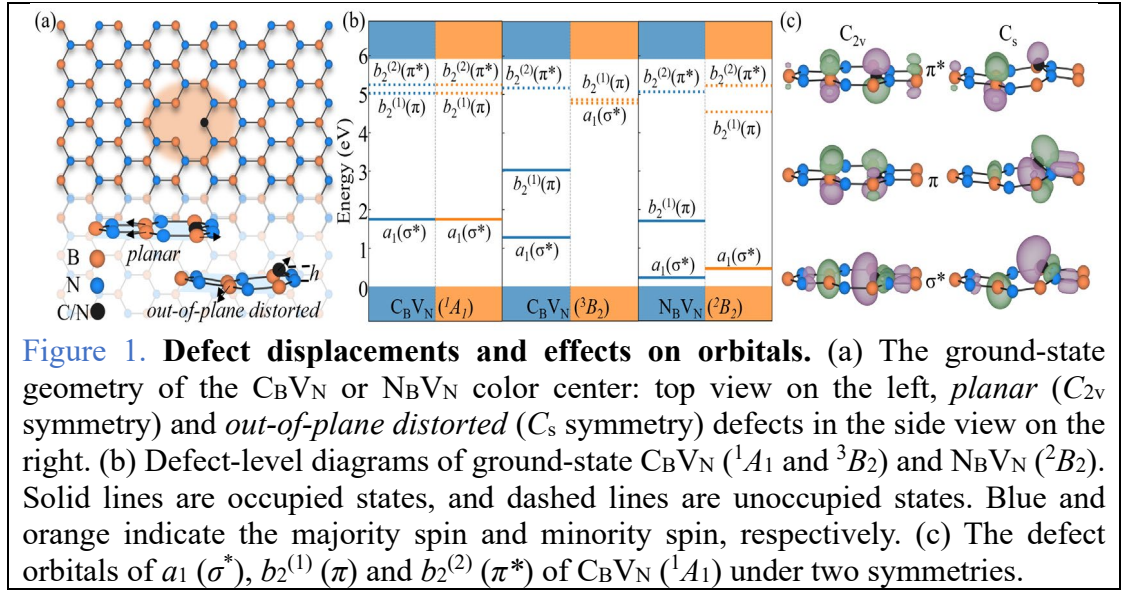

Figure 1. **Defect displacements and effects on orbitals.** (a) The ground-state geometry of the $C_BV_N$ or $N_BV_N$ color center: top view on the left, *planar* ($C_{2v}$ symmetry) and *out-of-plane distorted* ($C_s$ symmetry) defects in the side view on the right. (b) Defect-level diagrams of ground-state $C_BV_N$ ($^1A_1$ and $^3B_2$) and $N_BV_N$ ($^2B_2$). Solid lines are occupied states, and dashed lines are unoccupied states. Blue and orange indicate the majority spin and minority spin, respectively. (c) The defect orbitals of $a_1$ ($\sigma^*$), $b_2^{(1)}$ ($\pi$) and $b_2^{(2)}$ ($\pi^*$) of $C_BV_N$ ($^1A_1$) under two symmetries.

**RESULTS AND DISCUSSION**

*Out-of-plane displacement and impacts on defect states*

Figure 1a depicts $C_BV_N$ ($N_BV_N$) color center structures within the *h*-BN monolayer, which contain one nitrogen vacancy adjacent to carbon (nitrogen) substitution for boron. Their *planar* structures feature a $C_{2v}$ point group symmetry, which has been widely adopted by most previous theoretical studies [21,23,29]. However, recent studies suggest that anti-site carbon or nitrogen can be *out-of-plane distorted*, presenting a reduced $C_s$ symmetry [24,25]. This symmetry reduction can significantly impact the optical properties, as we discuss in later sections. Our HSE hybrid functional calculation predicts a 5.92 eV band gap of pristine *h*-BN monolayer, which is in agreement with the experimental value of approximately 6.0 eV [30-32]. Within the band gap, we identified three kinds of defect states, labeled as $a_1$, $b_2^{(1)}$, and $b_2^{(2)}$, following the irreducible representations of $C_{2v}$ symmetry. To simplify the discussion, we retain these $C_{2v}$ notations in $C_s$-symmetry structures. For the same reason, the multi-electron representations of spin-singlet $C_BV_N$, spin-triplet $C_BV_N$, and spin-doublet $N_BV_N$ are designated $^1A_1$, $^3B_2$, and $^2B_2$, respectively [33]. Figure 1b displayed the corresponding defect-level diagrams, with relative energies summarized in Table I. Figure 1c shows the real-space wavefunction of the defect states. The $a_1$ state appears as a $\sigma^*$-like antibonding state involving boron atoms proximal to the vacancy center and the anti-site atom (carbon or nitrogen), while the $b_2$ states, which are mainly derived by $p_z$ orbitals of the two neighboring boron atoms and the anti-site atom, reveal $\pi$- or $\pi^*$-like bonding characteristics.

Table I. HSE06 calculated relative energies of majority-spin defect levels with respect to the valence band maximum. Values in parentheses are minority spin states. The height of out-of-plane displacement ($h$, in Å) is also provided, along with the energies (in eV) involved in the optical process.

| System | State | Symmetry | $a_1$ | $b_2^{(1)}$ | $b_2^{(2)}$ | $h$ | $\Delta E_0$ | $\lambda_{GS}$ | $\lambda_{ES}$ |
|---|---|---|---|---|---|---|---|---|---|
| $C_BV_N$ | $^1A_1$ | $C_s$ | 2.63 (2.63) | 4.44 (4.44) | 5.83 (5.83) | 0.61 | 1.83 | 0.890 | 0.783 |
| | $^3B_2$ | $C_{2v}$ | 1.28 (4.78) | 3.03 (4.86) | 5.17 (5.96) | 0 | 1.96 | 0.307 | 0.292 |
| $N_BV_N$ | $^2B_2$ | $C_s$ | 0.70 (0.94) | 1.91 (4.36) | 5.19 (5.57) | 0.49 | 1.74 | 0.910 | 0.949 |

The ground states of these three defects partially or fully occupy the $\sigma^*$-like states $a_1$, where the repulsive nature of these antibonding states may provoke out-of-plane instability. If the atoms are restricted on the $h$-BN plane, the in-plane displacement of atoms can lower total energy by 1.4 eV, 0.7 eV, and 1.6 eV for $C_BV_N$ ($^1A_1$), $C_BV_N$ ($^3B_2$), and $N_BV_N$ ($^2B_2$), respectively (see Table SI). Beyond the planer assumption, allowing out-of-plane degrees of freedom can further lower the total energy. For example, the out-of-plane displacement for $C_BV_N$ ($^1A_1$), as shown in Figure 1c, effectively reduces the antibonding overlap of the $\sigma^*$-like bond, lowering total energy by 0.57 eV. This *out-of-plane distortion* thus forms the true ground state structure, and its *planar* structure is metastable on a saddle point. This agrees with our vibration calculation that reveals a pronounced -14.7 THz out-of-plane soft mode of *planar* $C_BV_N$ ($^1A_1$). In contrast, occupation of $\pi$-like defect state ($b_2^{(1)}$) tends to maintain the overlap of bonding orbitals thus stabilizing *planar* structures. For example, the ground state structure is *planar* in $C_BV_N$ ($^3B_2$) due to the cancelation of simultaneously occupying one antibonding $\sigma^*$-like state and one bonding $\pi$-like state. In $N_BV_N$ ($^2B_2$), with two $\sigma^*$-like and one $\pi$-like state occupied, its *planar* structure presents a -1.3 THz soft phonon mode, which is substantially reduced compared to $C_BV_N$ ($^1A_1$). This indicates that the additional $\sigma^*$-like state tilts the energy balance back toward the *out-of-plane distorted* structure.

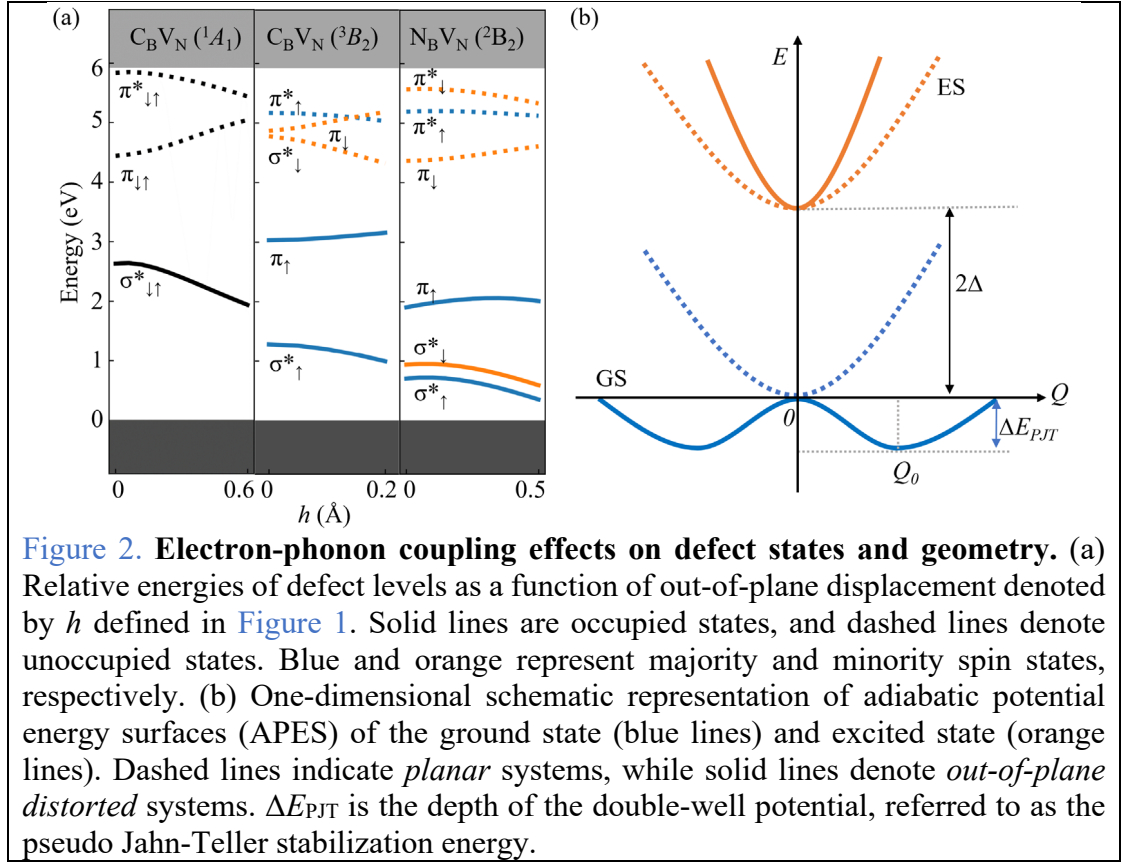

**Figure 2. Electron-phonon coupling effects on defect states and geometry.** (a) Relative energies of defect levels as a function of out-of-plane displacement denoted by *h* defined in Figure 1. Solid lines are occupied states, and dashed lines denote unoccupied states. Blue and orange represent majority and minority spin states, respectively. (b) One-dimensional schematic representation of adiabatic potential energy surfaces (APES) of the ground state (blue lines) and excited state (orange lines). Dashed lines indicate *planar* systems, while solid lines denote *out-of-plane distorted* systems. $\Delta E_{PJT}$ is the depth of the double-well potential, referred to as the pseudo Jahn-Teller stabilization energy.

To further explore the competition between $b_2^{(1)}$ and $a_1$ states, we evaluated the relative energies of defect states as a function of out-of-plane displacement in Figure 2a. For simplicity, the height (*h*) of the anti-site atom (carbon or nitrogen) relative to the plane was used to represent distortion amplitude. As expected, $C_BV_N$ ($^1A_1$) shows greater out-of-plane displacement (0.61 Å) compared to $N_BV_N$ ($^2B_2$, 0.49 Å). An assumed *h* of 0.2 Å for *planar* $C_BV_N$ ($^3B_2$) was also calculated for comparison. Across all three systems, with increasing displacement *h*, the relative energies of $a_1$ and $b_2^{(1)}$ states continually decrease and increase, respectively, which fully aligns with chemical bonds analysis. Specifically, the occupation of $\sigma^*$-like states favors out-of-plane distortion, while the occupation of $\pi$-like states helps maintain planar structures.

*Impact of overall vibronic coupling on geometry*

Beyond individual state effects, out-of-plane displacement and symmetry breaking in layered materials are fundamentally a product of structural nonadiabaticity derived from electron-phonon coupling [34]. This nonadiabatic effect represents an inherently infinite-state coupling problem within solid-state color centers. To encapsulate the collective geometric impacts, we applied an effective two-state PJT model [35]:

$$\epsilon(Q) = \frac{1}{2}K_0 Q^2 \pm (\Delta^2 + F^2 Q^2)^{\frac{1}{2}} + \Delta. \quad (1)$$

where $\Delta$ is half of the total energy difference between ground and the first excited states (illustrated in Figure 2b), $K_0$ is the primary force constant for the effective vibrational mode $\omega$ with a corresponding mass *M* along normal coordinate *Q*, given by $K_0 = M\omega^2$.

$F$ is the vibronic coupling between ground and excited states, resulting from electron-phonon coupling. The curvatures of $\epsilon(Q)$ at $Q=0$ for the ground and excited states are $K_l = K_0 \pm F^2/\Delta$, revealing that the PJT effect always softens the ground state and hardens the excited state. When $F^2/K_0 > \Delta$, known as the PJT instability condition, $K_l$ for the ground state becomes negative, leading to the spontaneously broken symmetry. It is worth mentioning that $F$ becomes nonzero only when the vibrational displacements $Q_\Gamma$ along normal coordinates are Jahn-Teller (JT) active. In the systems discussed in this work, the out-of-plane displacements predominantly derive from one quasi-local vibrational mode $Q_{B2}$. The symmetry of this mode enables strong vibronic coupling between the ground and first excited states of $C_BV_N$ ($^1A_1$) and $N_BV_N$ ($^2B_2$), where multi-electron representations of the first excited state are $^1B_2$ and $^2A_1$, respectively. In contrast, the first excited state of $C_BV_N$ ($^3B_2$) remains $^3B_2$, for which $Q_{B2}$ is no longer JT active, hence vibronic coupling $F$ is very small.

Table II. PJT parameters: primary force constant $K_0$ ($\hbar\omega$), vibronic coupling constant $F$, half the energy splitting between ground and excited states $\Delta$, PJT stabilization energy $\Delta E_{PJT}$, and displacement amplitude $Q_0$ ($\sqrt{amu}Å$). $K_0$, $F$, $\Delta$, $\Delta E_{PJT}$ are in units of meV.

| System | State | $K_0$ | $F$ | $\Delta$ | $Q_0$ | $\Delta E_{PJT}$ |
|---|---|---|---|---|---|---|
| $C_BV_N$ | $^1A_1$ | 31.7 | 259 | 647 | 2.97 | 537 |
|  | $^3B_2$ | 49.7 | 0.8 | 982 | 0 | 0 |
| $N_BV_N$ | $^2B_2$ | 28.2 | 189 | 833 | 1.97 | 74 |

PJT model fitting parameters are summarized in Table II by matching the double-well adiabatic potential energy surfaces (APES) shown in Figure 2b. Instead of selecting specific JT active $Q_\Gamma$ normal modes, a coordinate $Q$ is defined by a mass-weighted displacement vector between the high-symmetry *planar* ($r_{h;\alpha}$) and low-symmetry *out-of-plane distorted* structures ($r_{l;\alpha}$):

$$Q = \sum_{\alpha i} \sqrt{m_\alpha}(r_{h;\alpha i} - r_{l;\alpha i}), \qquad (2)$$

where $m_\alpha$ is the mass of atom $\alpha$, and $i$ ranges across the dimensions $x$, $y$, and $z$. Since only the first excited states are considered, we have $2\Delta = \Delta E_0$ (as defined in Figure 4). Our fitting parameters of $N_BV_N$ are consistent with the findings of a previous study [36]. As noted earlier, the out-of-plane displacement of $C_BV_N$ ($^1A_1$) is more pronounced than that of $N_BV_N$ ($^2B_2$), with the PJT stabilization energy of the former being over 7 times larger, an overall displacement $Q_0$ that is 50% larger, and vibronic coupling $F$ that is 37% greater, compared to those of the latter. In contrast, owing to symmetry constraints, the vibronic coupling of $C_BV_N$ ($^3B_2$) is only 0.8 meV, significantly below the 221 meV threshold necessary for PJT instability.

*Local vibrational modes with distinct features*

The vibrations of color centers comprise a series of quasi-local modes, vibrating in an almost molecular manner, while also coupling to fully delocalized bulk phonons. The degree of localization of each phonon can be quantified by the participation ratio (PR) [37]:

$$PR_k = \sum_{\alpha}\left(\sum_i \Delta r_{k;\alpha i}^2\right)^2, \quad (3)$$

where $\Delta r_{k;\alpha i}$ is the component of the normalized eigenvector of the $k$th vibrational mode for atom $\alpha$ along the $i$ (ranges $x$, $y$, and $z$) direction. A high PR means that only a small number of atoms highly participate in this phonon, indicating a quasi-local mode. Figure 3a displays the phonon density of states (PDOS) of pristine $h$-BN monolayer, including the projection onto the in-plane and out-of-plane displacements of B and N, along with a local vibrational analysis for the three defects. The phonon calculations are based on their ground state geometries, specifically, $C_s$ symmetry for $C_BV_N$ ($^1A_1$) and $N_BV_N$ ($^2B_2$) and $C_{2v}$ symmetry for $C_BV_N$ ($^3B_2$), resulting in distinct irreducible representations of the phonons.

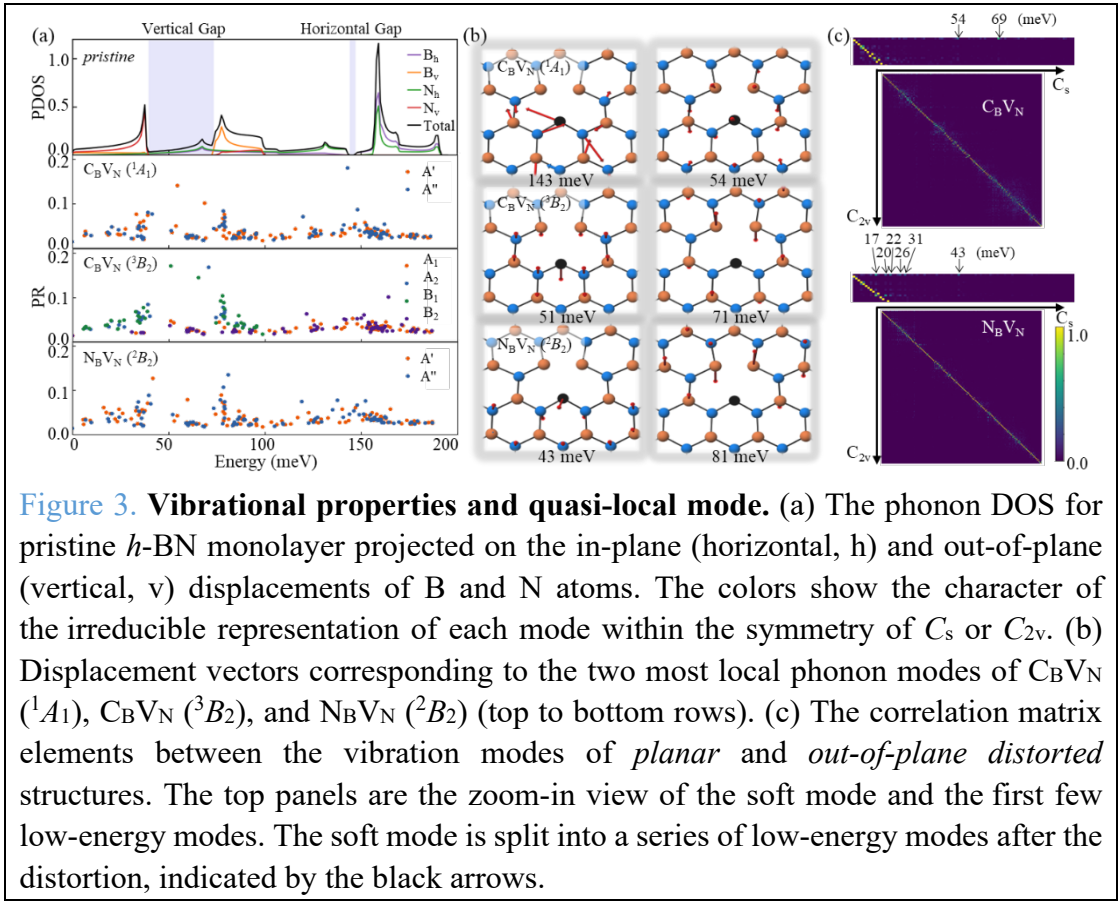

Figure 3. **Vibrational properties and quasi-local mode.** (a) The phonon DOS for pristine $h$-BN monolayer projected on the in-plane (horizontal, h) and out-of-plane (vertical, v) displacements of B and N atoms. The colors show the character of the irreducible representation of each mode within the symmetry of $C_s$ or $C_{2v}$. (b) Displacement vectors corresponding to the two most local phonon modes of $C_BV_N$ ($^1A_1$), $C_BV_N$ ($^3B_2$), and $N_BV_N$ ($^2B_2$) (top to bottom rows). (c) The correlation matrix elements between the vibration modes of *planar* and *out-of-plane distorted* structures. The top panels are the zoom-in view of the soft mode and the first few low-energy modes. The soft mode is split into a series of low-energy modes after the distortion, indicated by the black arrows.

In the pristine monolayer, a vertical gap appears in the PDOS in a low energy region (40-75 meV), where all atoms exhibit purely in-plane vibrations. A smaller horizontal gap between 140-149 meV exists at higher energies. Mapping PR scatter plots onto the pristine PDOS reveals that most phonons with low PRs are bulk modes, while the remaining are quasi-local modes. As shown in Figure 3b, the top two local modes of *out-of-plane distorted* ground states include combined out-of-plane and in-plane components, in stark contrast to those of the *planar* structures. More specifically, lower energy quasi-local modes near the vertical gap predominantly comprise out-of-plane displacements, while those near the horizontal gap are mainly in-plane. These out-of-plane quasi-local modes near the vertical gap could serve as fingerprints to facilitate the experimental identification of specific defect types. To further support experiments, we confirmed that quasi-local phonons are both Raman and IR active, except the 71 meV mode of $C_BV_N$ ($^3B_2$), which is solely Raman active.

To analyze vibrational state variations caused by out-of-plane distortions, in Figure 3c, we compared eigenvectors pairwise using a correlation function:

$$\text{cor}(k, k') = \left| \sum_{i,\alpha} \Delta r_{k;\alpha i} \Delta r_{k';\alpha i} \right|. \quad (4)$$

Correlation values range from 0 to 1, where 0 and 1 denote that the $k$ and $k'$ modes are orthogonal and identical pairs, respectively. Most non-zero elements lie along the diagonal of the correlation matrix, with values close to 1, indicating that most phonon modes remain unaltered by the out-of-plane displacement, with minimal perturbation in phonon polarization vectors as anticipated for bulk phonons. However, some phonons exhibit major polarization changes after the out-of-plane distortions, correlating with other phonon groups associated with the defects. In particular, the soft modes of the *planar* $C_BV_N$ ($^1A_1$) and $N_BV_N$ ($^2B_2$), which consist mostly of out-of-plane displacement of the anti-site carbon or nitrogen atom, lose their pure out-of-plane character after distortion. Correlation analysis reveals these soft modes do not simply shift to higher energies, but split into multiple local modes upon distortion. The $C_BV_N$ soft mode mainly splits into 54 meV and 69 meV modes, aligning with the two most localized vibration modes near the vertical gap of pristine monolayer h-BN, while the $N_BV_N$ soft mode decomposes into a broader range of low-energy modes.

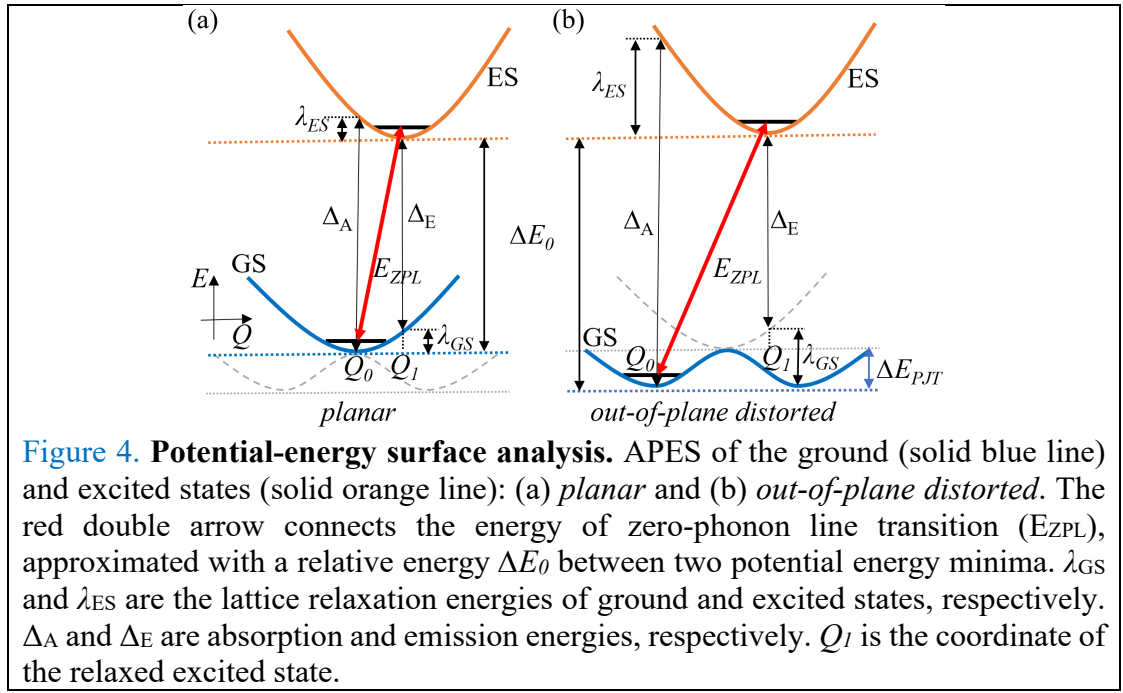

Figure 4. **Potential-energy surface analysis.** APES of the ground (solid blue line) and excited states (solid orange line): (a) *planar* and (b) *out-of-plane distorted*. The red double arrow connects the energy of zero-phonon line transition ($E_{ZPL}$), approximated with a relative energy $\Delta E_0$ between two potential energy minima. $\lambda_{GS}$ and $\lambda_{ES}$ are the lattice relaxation energies of ground and excited states, respectively. $\Delta_A$ and $\Delta_E$ are absorption and emission energies, respectively. $Q_1$ is the coordinate of the relaxed excited state.

*Dominating effects on quantum emission*

Quantum emission involves simultaneous electronic and vibrational transitions. In the classical Franck-Condon picture illustrated in Figure 4, two types of optical transition loops exist in these systems, each encompassing four steps: (i) optical excitation from the ground state to excited state at geometry $Q_0$ with energy of $\Delta_A$; (ii) structural relaxation in the excited state from $Q_0$ to $Q_1$ with energy of $\lambda_{ES}$; (iii) optical emission from excited to ground state at geometry $Q_1$ with energy of $\Delta_E$; (iv) ground state structural relaxation from $Q_1$ back to $Q_0$ with an energy of $\lambda_{GS}$.

Regarding the potential energy surfaces, *planar* and *out-of-plane distorted* systems possess single-well and double-well ground state potentials, respectively. Their excited states, however, are all single-well potentials, which agrees with the PJT picture. This difference yields several significant effects. First, *out-of-plane distorted* systems tend to have larger $\Delta_A$ due to the additional stabilization energy $\Delta E_{PJT}$. On the contrary, $\Delta_E$ remains similar for both types of geometries, since they share identical single-well ground and excited state potentials. Second, *out-of-plane distorted* systems experience considerably larger structural relaxation energies $\lambda_{GS}$ and $\lambda_{ES}$ with larger optical transition displacement vector, $\Delta Q$, between excited and ground state minima ($\Delta Q = Q_1 - Q_0$). As a result, greater $\Delta Q$ invokes a greater contribution from vibrational modes to the emission process, which engenders a more pronounced PSB in the PL spectrum.

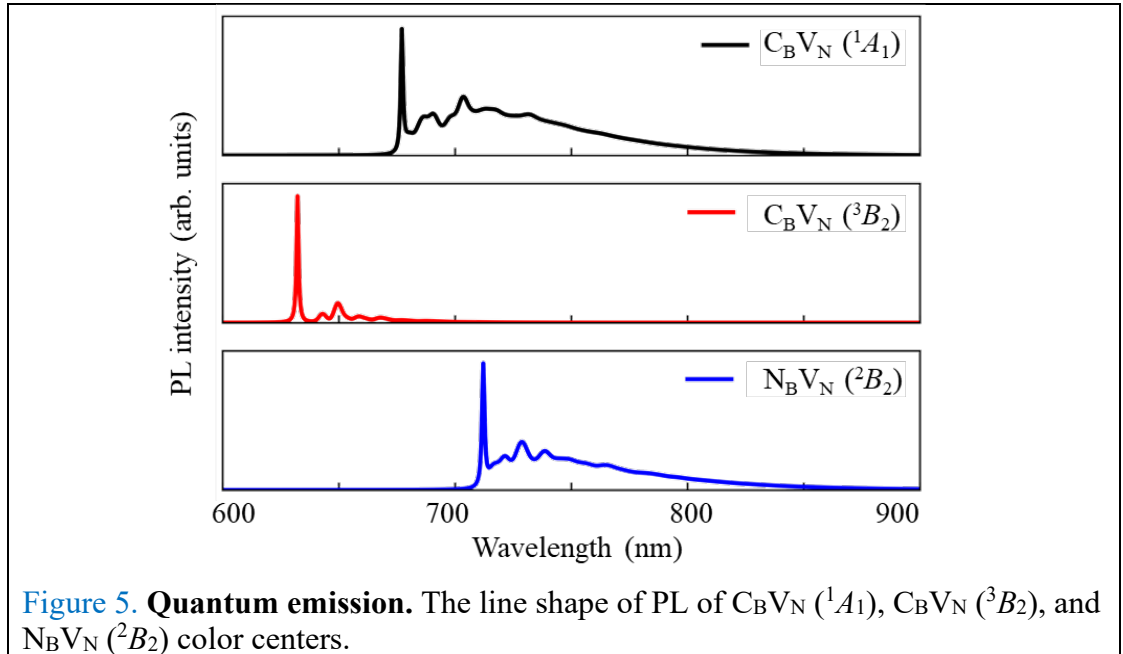

Figure 5. **Quantum emission.** The line shape of PL of $C_BV_N$ ($^1A_1$), $C_BV_N$ ($^3B_2$), and $N_BV_N$ ($^2B_2$) color centers.

To validate these assumptions, we calculated the normalized PL line shapes for $C_BV_N$ ($^1A_1$), $C_BV_N$ ($^3B_2$), and $N_BV_N$ ($^2B_2$), presented in Figure 5. The ZPL peaks, approximated by $\Delta E_0$, are at 1.83 eV (677 nm), 1.96 eV (633 nm), and 1.74 eV (712 nm), respectively. As anticipated, *out-of-plane distorted* $C_BV_N$ ($^1A_1$) and $N_BV_N$ ($^2B_2$) exhibit substantial PSB contributions, similar to 'Group 1' emitters found in previous experiments [17]. More specifically, the $C_BV_N$ ($^1A_1$) PSB spans a broader energy range with a considerably larger relative area than $N_BV_N$ ($^2B_2$), consistent with its larger out-of-plane displacement, larger $\Delta Q$ during optical transitions, as well as larger relaxation energies $\lambda_{GS}$ and $\lambda_{ES}$. On the contrary, *planar* $C_BV_N$ ($^3B_2$) resembles an ideal single photon emitter (akin to the 'Group 2' emitters experimentally [17]), characterized by a dominant ZPL peak and negligible PSB, which better protects photon coherence. This further suggests that color centers with higher symmetry in layered materials should be more desirable for quantum applications.

**CONCLUSION**

In summary, we conducted a comprehensive investigation of the electronic and vibrational properties of $C_BV_N$ and $N_BV_N$ color centers to elucidate the primary structural factors governing the performance of single-photon emitters. Their ground-state geometries could be *planar* ($C_BV_N$ $^3B_2$) or *out-of-plane distorted* ($C_BV_N$ $^1A_1$ and $N_BV_N$ $^2B_2$), determined by the occupations of defect states. We discussed the origin of out-of-plane displacement by examining defect state bonding/antibonding natures, where general geometry effects can be understood from PJT instability. Local vibrational analysis revealed a series of distinct quasi-local phonons that could serve as fingerprints to facilitate experimental identification of specific point defect types. Furthermore, we connected these structural insights to the contrast in PL line shapes between *planar* and *out-of-plane distorted* defects. These insights suggest broader applicability to layered materials and will accelerate the design of optimal quantum emitters.

**METHODS**

DFT calculations [38,39] were performed using the Vienna ab initio simulation package (VASP) [40,41] with projected augmented wave (PAW) pseudopotentials [42,43]. The exchange-correlation functional was within generalized gradient approximations (GGA) in the form of Perdew-Burke-Ernzerhof (PBE) [44]. The energy cutoff of the plane wave basis was set to 600 eV. A 30 Å vacuum layer was added to eliminate the interlayer interaction between periodic images. The PBE optimized in-plane lattice constant $a$ of monolayer $h$-BN unit cell was 2.512 Å, held fixed during the supercell calculations. Color centers were built in 6×6×1 supercells, with a 3×3×1 $k$-grid equivalent to an 18×18×1 supercell of the primitive cell. Structures were relaxed until the total energy converged within $10^{-7}$ eV. The specific spin configurations with neutral charge states were considered by fixing state occupations. The HSE06 functional is used to calculate the electronic properties [45]. Vibration modes were obtained by the finite displacement method [46,47] as implemented in PHONOPY [48]. PL line shapes were computed using the Huang-Rhys approximation [28,49], with further implementation details provided in Ref [50,51].

**ACKNOWLEDGMENTS**

Work at UB was supported by the National Science Foundation under Grant No. DMREF-1626967. Work at UT was supported by the Welch Foundation under grant F-2094. Work at HDU was supported by Zhejiang Provincial Natural Science Foundation (QN25A040026). Work at SUSTech was supported by National Natural Science Foundation of China (12104207). Computational resources and support were provided by the Center for Computational Research, University at Buffalo, SUNY, and the Texas Advanced Computing Center (TACC).

**AUTHOR CONTRIBUTIONS**

Z.T. and F.J. carried out the DFT calculations under the supervision of P.Z. All authors contributed to the discussion and writing the manuscript. P.Z. conceived and led the entire scientific project.

**COMPETING INTERESTS**

The authors declare no competing interests.

**REFERENCE**

[1]    T. T. Tran, K. Bray, M. J. Ford, M. Toth, and I. Aharonovich, Nature Nanotechnology **11**, 37 (2016).
[2]    S. Ren, Q. Tan, and J. Zhang, Journal of Semiconductors **40**, 071903 (2019).
[3]    S. I. Azzam, K. Parto, and G. Moody, Applied Physics Letters **118** (2021).
[4]    I. Aharonovich, D. Englund, and M. Toth, Nature Photonics **10**, 631 (2016).
[5]    X. Liu and M. C. Hersam, Nature Reviews Materials **4**, 669 (2019).
[6]    G. Grosso *et al.*, Nature Communications **8**, 705 (2017).
[7]    K. Yamamura, N. Coste, H. Z. J. Zeng, M. Toth, M. Kianinia, and I. Aharonovich, Nanophotonics  (2024).
[8]    S. Vaidya, X. Gao, S. Dikshit, I. Aharonovich, and T. Li, Advances in Physics: X **8**, 2206049 (2023).
[9]    A. Gottscholl *et al.*, Nature Communications **12**, 4480 (2021).
[10]   A. Cobarrubia, N. Schottle, D. Suliman, S. Gomez-Barron, C. R. Patino, B. Kiefer, and S. K. Behura, ACS Nano **18**, 22609 (2024).
[11]   E. Glushkov *et al.*, ACS Nano **16**, 3695 (2022).
[12]   A. Gale, C. Li, Y. Chen, K. Watanabe, T. Taniguchi, I. Aharonovich, and M. Toth, ACS Photonics **9**, 2170 (2022).
[13]   A. Gottscholl *et al.*, Nature materials **19**, 540 (2020).
[14]   R. Bourrellier, S. Meuret, A. Tararan, O. Stéphan, M. Kociak, L. H. G. Tizei, and A. Zobelli, Nano Letters **16**, 4317 (2016).
[15]   T. T. Tran *et al.*, ACS nano **10**, 7331 (2016).
[16]   X. Li, G. D. Shepard, A. Cupo, N. Camporeale, K. Shayan, Y. Luo, V. Meunier, and S. Strauf, ACS nano **11**, 6652 (2017).
[17]   A. Sajid, M. J. Ford, and J. R. Reimers, Reports on Progress in Physics **83**, 044501 (2020).
[18]   J. C. Meyer, A. Chuvilin, G. Algara-Siller, J. Biskupek, and U. Kaiser, Nano Letters **9**, 2683 (2009).
[19]   C. Jin, F. Lin, K. Suenaga, and S. Iijima, Physical Review Letters **102**, 195505 (2009).
[20]   C. Attaccalite, M. Bockstedte, A. Marini, A. Rubio, and L. Wirtz, Physical Review B **83**, 144115 (2011).
[21]   S. A. Tawfik, S. Ali, M. Fronzi, M. Kianinia, T. T. Tran, C. Stampfl, I. Aharonovich, M. Toth, and M. J. Ford, Nanoscale **9**, 13575 (2017).
[22]   G. Cheng, Y. Zhang, L. Yan, H. Huang, Q. Huang, Y. Song, Y. Chen, and Z. Tang, Computational Materials Science **129**, 247 (2017).
[23]   A. Sajid, J. R. Reimers, and M. J. Ford, Physical Review B **97**, 064101 (2018).

[24]  G. Noh, D. Choi, J.-H. Kim, D.-G. Im, Y.-H. Kim, H. Seo, and J. Lee, Nano Letters **18**, 4710 (2018).
[25]  S. Li, J.-P. Chou, A. Hu, M. B. Plenio, P. Udvarhelyi, G. Thiering, M. Abdi, and A. Gali, npj Quantum Information **6**, 85 (2020).
[26]  M. E. Turiansky and C. G. Van de Walle, 2D Materials **8**, 024002 (2021).
[27]  S. Gao, H.-Y. Chen, and M. Bernardi, npj Computational Materials **7**, 85 (2021).
[28]  K. Huang and A. Rhys, Proceedings of the Royal Society of London. Series A. Mathematical and Physical Sciences **204**, 406 (1950).
[29]  G. D. Cheng, Y. G. Zhang, L. Yan, H. F. Huang, Q. Huang, Y. X. Song, Y. Chen, and Z. Tang, Computational Materials Science **129**, 247 (2017).
[30]  Y. Shi *et al.*, Nano Letters **10**, 4134 (2010).
[31]  C. Zhi, Y. Bando, C. Tang, H. Kuwahara, and D. Golberg, Advanced Materials **21**, 2889 (2009).
[32]  J. Sun *et al.*, Chemical Society Reviews **47**, 4242 (2018).
[33]  M. Abdi, J.-P. Chou, A. Gali, and M. B. Plenio, ACS Photonics **5**, 1967 (2018).
[34]  I. B. Bersuker, Chemical Reviews **121**, 1463 (2020).
[35]  I. Bersuker, *The Jahn-Teller effect and vibronic interactions in modern chemistry* (Springer Science & Business Media, 2013).
[36]  S. Li, J.-P. Chou, A. Hu, M. B. Plenio, P. Udvarhelyi, G. Thiering, M. Abdi, and A. Gali, npj Quantum Information **6**, 1 (2020).
[37]  R. J. Bell, P. Dean, and D. C. Hibbins-Butler, Journal of Physics C: Solid State Physics **3**, 2111 (1970).
[38]  W. Kohn and L. J. Sham, Physical Review **140**, A1133 (1965).
[39]  P. Hohenberg and W. Kohn, Physical Review **136**, B864 (1964).
[40]  G. Kresse and J. Furthmüller, Computational Materials Science **6**, 15 (1996).
[41]  G. Kresse and J. Furthmüller, Physical Review B **54**, 11169 (1996).
[42]  P. E. Blöchl, Physical Review B **50**, 17953 (1994).
[43]  G. Kresse and D. Joubert, Physical Review B **59**, 1758 (1999).
[44]  J. P. Perdew, K. Burke, and M. Ernzerhof, Physical Review Letters **77**, 3865 (1996).
[45]  A. V. Krukau, O. A. Vydrov, A. F. Izmaylov, and G. E. Scuseria, The Journal of Chemical Physics **125**, 224106 (2006).
[46]  G. Kresse, J. Furthmüller, and J. Hafner, Europhysics Letters (EPL) **32**, 729 (1995).
[47]  K. Parlinski, Z. Q. Li, and Y. Kawazoe, Physical Review Letters **78**, 4063 (1997).
[48]  A. Togo and I. Tanaka, Scripta Materialia **108**, 1 (2015).
[49]  L. Razinkovas, M. W. Doherty, N. B. Manson, C. G. Van de Walle, and A. Alkauskas, Physical Review B **104**, 045303 (2021).
[50]  A. Alkauskas, B. B. Buckley, D. D. Awschalom, and C. G. Van de Walle, New Journal of Physics **16**, 073026 (2014).
[51]  S. A. Tawfik and S. P. Russo, Computer Physics Communications **273**, 108222 (2022).

# SUPPLEMENTAL MATERIAL

Table SI. The relative energies of $C_BV_N$ ($^1A_1$ and $^3B_2$) and $N_BV_N$ ($^2B_2$) with or without in-plane or out-of-plane displacements. $\Delta E_{PJT}$ and $h$ are PJT stabilization energy and height of out-of-plane displacement, respectively.

| System | State | Structure | Relative energy (eV) | $\Delta E_{PJT}$ (meV) | $h$ (Å) |
|---|---|---|---|---|---|
| $C_BV_N$ | $^1A_1$ | unrelaxed | 0 | | 0 |
| | | in-plane distorted | -1.394 | | 0 |
| | | out-of-plane distorted | -0.625 | | 0.61 |
| | | fully relaxed | -1.931 | 537 | 0.61 |
| | $^3B_2$ | unrelaxed | -0.564 | | 0 |
| | | fully relaxed | -1.231 | | 0 |
| $N_BV_N$ | $^2B_2$ | unrelaxed | 0 | | 0 |
| | | in-plane distorted | -1.563 | | 0 |
| | | out-of-plane distorted | -0.378 | | 0.49 |
| | | fully relaxed | -1.637 | 74 | 0.49 |

Table SII. The relative energy (using HSE06, in unit of eV) of single-electron defect levels relative to the valence band. Values in parentheses are from PBE calculations.

| Structure | System | State | Majority spin | | | Minority spin | | |
|---|---|---|---|---|---|---|---|---|
| | | | $a_1$ | $b_2^{(1)}$ | $b_2^{(2)}$ | $a_1$ | $b_2^{(1)}$ | $b_2^{(2)}$ |
| planar | $C_BV_N$ | $^1A_1$ | 2.45 (2.62) | 4.58 (3.29) | 5.73 (4.40) | 2.45 (2.62) | 4.58 (3.29) | 5.73 (4.40) |
| | | $^3B_2$ | 1.28 (1.49) | 3.03 (2.59) | 5.16 (4.00) | 4.76 (3.22) | 4.86 (3.43) | |
| | $N_BV_N$ | $^2B_2$ | 0.49 (0.70) | 1.91 (1.92) | 5.10 (3.86) | 0.69 (0.87) | 4.29 (2.80) | 5.47 (4.06) |
| out-of-plane distorted | $C_BV_N$ | $^1A_1$ | 1.76 (1.70) | 5.03 (3.61) | 5.26 (3.97) | 1.76 (1.70) | 5.03 (3.61) | 5.26 (3.97) |
| | $N_BV_N$ | $^2B_2$ | 0.25 (0.40) | 1.71 (1.87) | 5.07 (3.79) | 0.46 (0.56) | 4.51 (2.97) | 5.22 (3.88) |